\documentclass[runningheads]{llncs}
\usepackage{graphicx}
\usepackage{booktabs}
\usepackage{multirow}
\usepackage{hyperref}
\usepackage{amssymb}
\usepackage{amsmath}
\usepackage{cleveref}
\usepackage{color}
\begin{document}
\title{XctDiff: Reconstruction of CT Images with Consistent Anatomical Structures from a Single Radiographic Projection Image}
\titlerunning{XctDiff}
%
\author{Qingze Bai\inst{1} \and Tiange Liu \inst{*,2} \and Zhi Liu \inst{1}
Yubing Tong\inst{3} \and
Drew Torigian\inst{3} \and
Jayaram Udupa\inst{3}
}

\authorrunning{Bai et al.}
\institute{Shandong University \and
The University of Science and Technology Beijing \and
The University of Pennsylvania \\
\email{liutiange623@126.com}
}
\maketitle              
\begin{abstract}
In this paper, we present XctDiff, an algorithm framework for reconstructing CT from a single radiograph, which decomposes the reconstruction process into two easily controllable tasks: feature extraction and CT reconstruction. Specifically, we first design a progressive feature extraction strategy that is able to extract robust 3D priors from radiograph. Then, we use the extracted prior information to guide the CT reconstruction in the latent space. Moreover, we design a homogeneous spatial codebook to improve the reconstruction quality further. The experimental results show that our proposed method achieves state-of-the-art reconstruction performance and overcome the blurring issue. We also apply XctDiff on self-supervised pre-training task. The effectiveness indicates that it has promising additional applications medical image analysis. The code is available at: \url{https://github.com/qingze-bai/XctDiff}
\end{abstract}
\section{Introduction}
Radiography and Computed Tomography (CT) are prevalent non-invasive imaging techniques in clinical medicine. They share similar imaging principles but have differences in application scenarios \cite{c:03,c:14}. Radiography, with low radiation exposure, is used for preliminary examinations like diagnosing bone fractures and pneumonia due to limited 3D information. In contrast, CT, with higher radiation exposure, can capture more intricate structures and lesions, making it suitable for diagnosing and treating complex diseases. Given these distinctions, an intriguing question arises: can radiograph be reconstructed as CT by deep learning and replace its function on certain tasks?

Theoretically, a radiograph is formed when detector receives X-rays that have been attenuated by the body and converts them into electrical signals. CT scan, on the other hand, obtains multiple X-ray projection data from different camera positions and then reconstructs them based on the Radon transform. Based on similar imaging principles, converting a CT volume to a radiograph can be considered as a relatively straightforward lossy compression process (e.g., digital radiography (DRR) technology), while reversing this process poses a challenging single-view reconstruction problem. 

Recently, some data-driven reconstruction algorithms have demonstrated the ability to generate 3D objects from single natural images, such as based on depth estimation \cite{c:22,c:27} or implicit representations \cite{c:20}. However, the differences in imaging principles between natural and medical images, as well as the requirement for internal structure modeling in medical images, result in single-view reconstruction in the medical image domain not benefiting from these methods. Some studies \cite{c:35} have used two orthogonal projections to reconstruct the CT image, but this requires specialized design and equipment. Others \cite{c:24,c:31} have employed Convolution Neural Network (CNN) or Generative Adversarial Network (GAN) to learn the mapping function from radiographs to CT scans. However, these methods suffer from severe image degradation and lack evaluation for potential applications.

\begin{figure}[t]
\includegraphics[width=\textwidth]{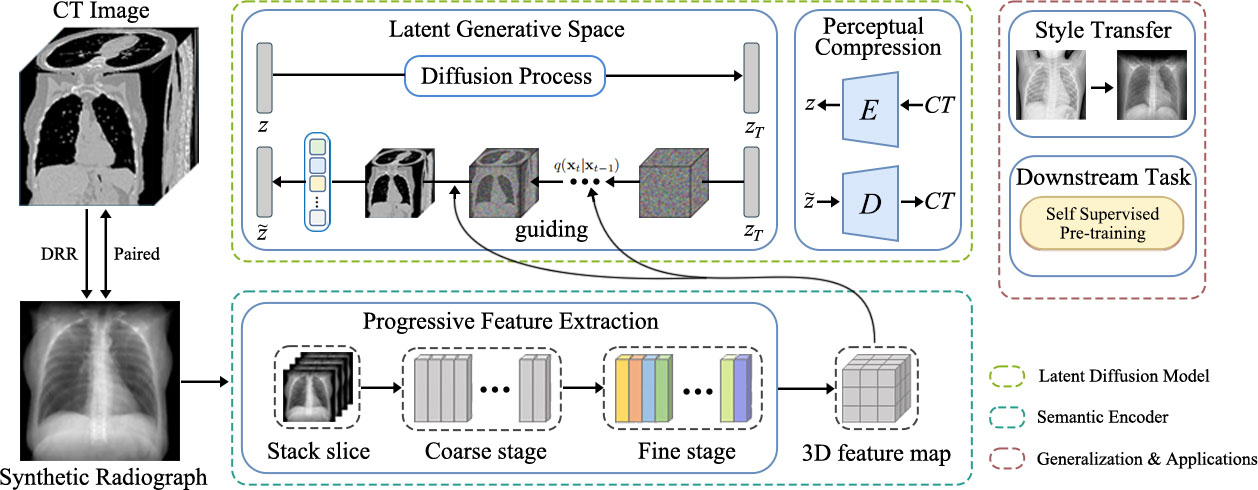}
\caption{The XctDiff utilizes a progressive semantic encoder to extract 3D anatomical priors from input radiograph image. The extracted features are then used to guide CT reconstruction in latent space. Finally, the reconstructed CT feature maps are used to generate high-quality CT images with consistent anatomical structures after passing through a vector quantization encoder. Note that the radiographs used in the training were converted using Digitally Reconstructed Radiography (DRR) technology. The radiographs in the inference stage are converted from real radiographs through style transfer model.} 
\label{figure01}
\end{figure}

In this paper, we propose a \textbf{radiograph to computed tomography image diffusion model (XctDiff)}, an algorithmic framework for CT image reconstruction from a single radiograph. As depicted in Fig.~\ref{figure01}, XctDiff consists of three main components: a perceptual compression model, a progressive semantic encoder for extracting 3D structural information from 2D radiograph, and a conditional diffusion model (DM). We decompose the reconstruction process into two parts: feature extraction and 3D reconstruction. Specifically, we extract anatomical information from radiograph and then utilize the extracted prior information to guide CT image reconstruction. We also proposed a homogeneous spatial codebook to enhance the reconstruction quality further. After all, all components are integrated and generalized for real radiographs. The reconstructed CT images are utilized on self-supervised pre-training task. Experimental results demonstrate that XctDiff achieves state-of-the-art reconstruction performance. 

\section{Method}

\subsection{3D Perceptual Compression Model}
Reconstructing CT images is equivalent to learning volume structure, which presents greater computational and pattern coverage challenges than 2D images. Inspired by latent generation models \cite{c:11,c:23}, we adopt a similar compression design to accommodate a 3D format input. For the architecture aspect, a homogeneous spatial codebook and an extra super-resolution module \cite{c:26} are incorporated to improve the reconstruction quality. For the loss function, we use 2D and 3D discriminators to optimize jointly the whole and slice.

Specifically, for given input CT $ x \in \mathbb{R}^{H \times W \times D} $, it is compressed by encoder $E$ into a latent representation $ z \in \mathbb{R}^{h \times w \times d \times n_{z} } $,  where $h, w$, and $d$ denote the feature map size in the latent space, $n_{z}$ represents the dimension of the codebook entries. Each 
spatial representation is then quantized element-by-element 
into the closest codebook entity, which can be expressed as:
\begin{equation}\label{eq1}
z_{q}=q(z):=\min_{z_{n} \in \mathcal{Z}}\left \| z_{ijk} - z_{n}     \right \|
\end{equation}
where $z_{n}$ and $n$ denote the codebook entity and the number of entities, respectively. $\mathcal{Z}$ is the vector quantization encoder.

To ensure perceptually rich variance space for generation, we adopt the heuristic strategy of vector quantization encoder \cite{c:11,c:36} and utilize z-score normalization to transform the variance space into a homogeneous space, which avoids the smoothing generated by the encoder against the codebook and thus improves the quality of reconstruction. The training objectives of the codebook can be summarized as follows:
\begin{equation}\label{eq2}
\begin{split}
    \mathcal{L}_{vq}=\left\| \mathcal{N} (sg[E(x)]) - \mathcal{N}(z_{q}) \right \|^{2}_{2} \\ +\left\|\mathcal{N}(sg[z_{q}])-\mathcal{N}(E(x)) \right \|^{2}_{2}
\end{split}
\end{equation}
where $\mathcal{N}$ represents the z-score normalization, $sg[\cdot]$ denotes the stop-gradient operation. We also introduce a super-resolution module at the end of decoder $G$ to improve the reconstruction performance. In addition, we apply a combination of slice discriminator $D_{2d}$ and volume discriminator $D_{3d}$ to optimize further jointly the autoencoder. The optimization objective of the discriminator can be formulated as:
\begin{equation}\label{eq3}
\begin{split}
    \mathcal{L}_{gan}=\alpha[ \log{D_{3d}(x)} + \log (1-D_{3d}(\hat{x}))] \\ + \beta[\log{D_{2d}(x_{k})} + \log (1-D_{2d}(\hat{x}_{k}))]
\end{split}
\end{equation}
where $x$ and $\hat{x}$ represent ground truth and prediction, respectively. $x_{k}$ denotes the $k th$ slice in the CT images. Finally, the optimization function of the autoencoder is summarized as follows:
\begin{equation}\label{eq4}
\mathcal{L}_{vqgan}=\lambda_{1} \mathcal{L}_{rec}+\lambda_{2} \mathcal{L}_{lpips}+\lambda_{3} \mathcal{L}_{vq}+\lambda_{4} \mathcal{L}_{gan}
\end{equation}
where $\mathcal{L}_{rec}$ and $\mathcal{L}_{lpips}$ denote L1 loss and perceptual loss, $\lambda_{i}$ are weighted hyperparameters.

\begin{figure}[t]
\centering
\includegraphics[width=\textwidth]{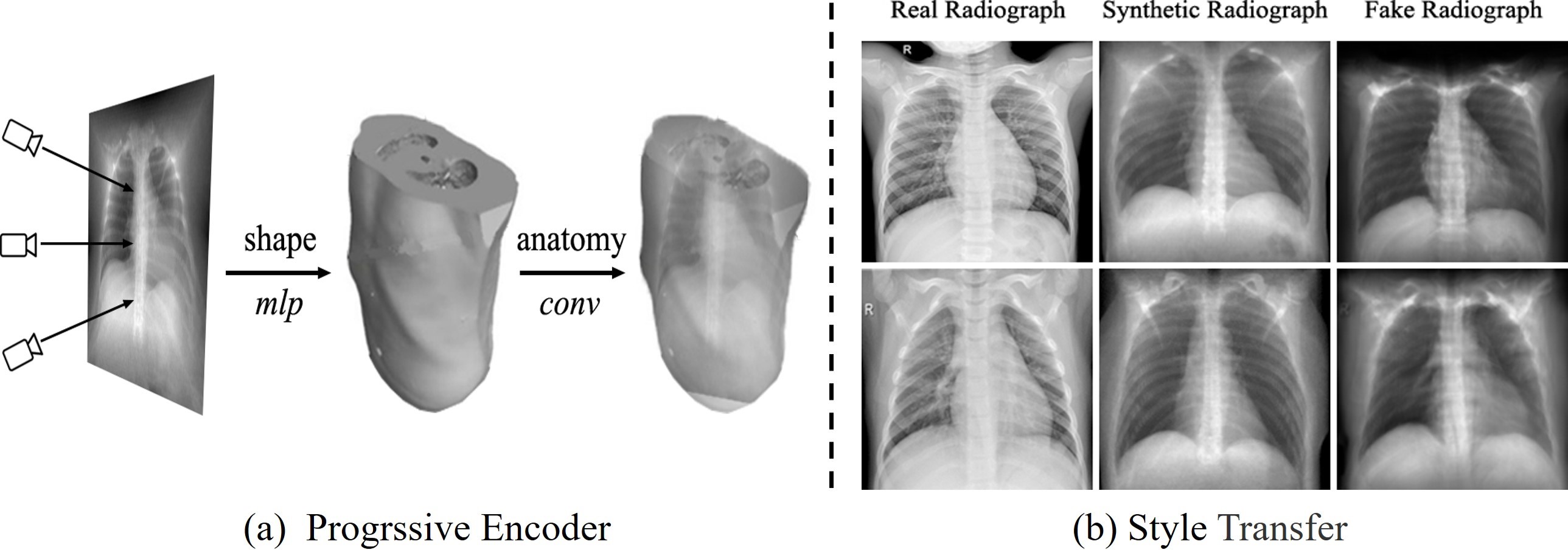}
\caption{(a) The progressive encoder firstly approximates the rough shape of the human body only in the coronal plane, then learns more accurate 3D anatomical representations through multiple successive convolutional layers. (b) Different styles of radiographs. (Left) Real radiographs from the ChestXray2017. (Middle) Synthesized radiographs used for training and evaluate. (Right) Synthesized style real radiographs.} 
\label{figure02}
\end{figure}
\subsection{Extract 3D Anatomical Information}
Reconstruction of a CT image from a single radiograph requires 3D prior information. We decouple the mapping relationship between the radiograph and the corresponding CT scan into two phases: shape mapping and anatomical structure mapping. As shown in Fig~\ref{figure02}(a), we proposed a progressive encoder $PE$ to obtain robust 3D priors by simplifying the complex mapping relationship. Specifically, for given an input radiograph $p$, we use multilayer perceptron (MLP) to convert the multilayer stacked slices into the human 3D space. The MLP only works in one direction since frontal projection radiograph matches in geometry a coronal slice, making it easier to learn the shape of the organs from this direction, which can be expressed as follows:
\begin{equation}\label{eq5}
\hat{v}_{ijk}= MLP(p_{ij}^{1},\cdot \cdot \cdot ,p_{ij}^{k})  
\end{equation}
where $v$ and $p$ denote the voxel of the CT image and the pixel of the radiograph, respectively. In the anatomical structure mapping stage, we globally optimize the extracted rough features by multiple successive 3D convolutional layers to obtain more accurate 3D anatomical priors. Finally, the extracted prior information is mapped into the homogeneous space by vector quantization encoder. Formally, this process can be summarized as:
\begin{equation}\label{eq6}
y= \mathcal{N}(q(Conv(\hat{v}))  
\end{equation}

\subsection{Prior Guided Diffusion Model}
The DM \cite{c:13,c:28,c:29} uses UNet $\theta$ to predict the added noise $\varepsilon$ at each time step $t$ in the inverse process. To minimize the pattern coverage challenge, we apply convolution only in the image plane (i.e., kernel size of 3×3×1). For given anatomical priors $y$ extracted from radiograph, the denoising encoder $\varepsilon_{\theta}(x,t,y)$ controls the direction of generation based on the prior distribution via cross-attention. The optimization function of the conditional diffusion model can be written as:
\begin{equation}\label{eq7}
\mathcal{L}_{dm}=E_{x,t,\varepsilon \sim \mathcal{N}(0,I)}(\left \|\varepsilon -\varepsilon_{\theta}(x_{t},t,y)\right \|_{2}^{2})
\end{equation}
where $x_{t}$ is the sampling of the diffusion process at the moment $t$, which is effected by noise schedule $\alpha_{t}$.

\section{Experiment}
\subsection{Datasets}
\label{sec:section4.1}
Similar to previous works \cite{c:24,c:31,c:35}, we used digitally reconstructed radiograph (DRR) technology to translate real CT scans into corresponding radiographs. However, as shown in Fig~\ref{figure02}(b), a disparity exists between synthetic radiographs and real ones in terms of realism and style, which may hinder model generalization and evaluation. Thus, when inferring real radiographs, we utilized CycleGAN \cite{c:41} to transform real radiographs into synthetic-style radiographs.

Experiments were conducted on four public datasets. The LIDC-IDRI dataset \cite{c:01}, comprising 1018 CT scan images, was employed for training and validation of our proposed framework, with voxel resolutions resampled to [2.5,2.5,2.5] and cropped to [128,128,128] cube regions. Among these, 916 cases were used for training and 102 cases were used for testing. In addition, the ChestXray2017 dataset \cite{c:10}, containing 5856 radiographic images, was used for style conversion between real and synthetic radiographic images. We randomly selected 1000 normal radiographs without lesions from this dataset for style migration and validation. For the BCV \cite{c:18} and MSD Spleen \cite{c:02} datasets, which are abdominal segmentation datasets, 25 cases were randomly selected using 5-fold cross-validation, with 5 and 7 cases were used for testing, respectively.

\subsection{Implementation Details}
\label{sec:section4.2}
All experiments were conducted on 2 RTX3090 GPUs. The XctDiff framework comprises two main training phases. Initially, we trained a perceptual compression model with a batch size of 1, a learning rate of 2e-4, and 80,000 iterations. Subsequently, a progressive encoder was trained with a batch size of 16, a learning rate of 1e-4, and 50,000 iterations. In the second stage, a conditional DM in latent space was trained with a batch size of 8, a learning rate of 1e-4, and 100,000 iterations. For the self-supervised pre-training task, we utilized the code library provided by Swin UNETR \cite{c:32}. Each model was cross-validated on the training dataset with 5 folds and trained for 50,000 iterations.

\begin{table}[b]
    \centering
    \caption{Evaluation of single-view CT reconstruction quality on the LIDC-IDRI dataset, where the mean (and Std) are reported.}
    \tabcolsep=4pt
    \begin{tabular}{l|ccc}
    \toprule
    & PSNR $\uparrow$ & SSIM $\uparrow$ & LPIPS $\downarrow$ \\
    \midrule
    ReconNet & 22.28($\pm1.237$) & 0.470($\pm0.068$) & 0.237($\pm0.048$)\\
    X2CTCNN  & 22.47($\pm1.460$) & 0.495($\pm0.077$) & 0.223($\pm0.056$) \\
    X2CTGAN  & 22.66($\pm1.442$) & 0.503($\pm0.074$) & 0.217($\pm0.070$) \\
    \midrule
    XctDiff  & \textbf{23.57($\pm$1.205)} & \textbf{0.524($\pm$0.075)} & \textbf{0.123($\pm$0.069)} \\
    \bottomrule
    \end{tabular}
\label{tab:table1}
\end{table}
\begin{figure}[t]
\centering
\includegraphics[width=\textwidth]{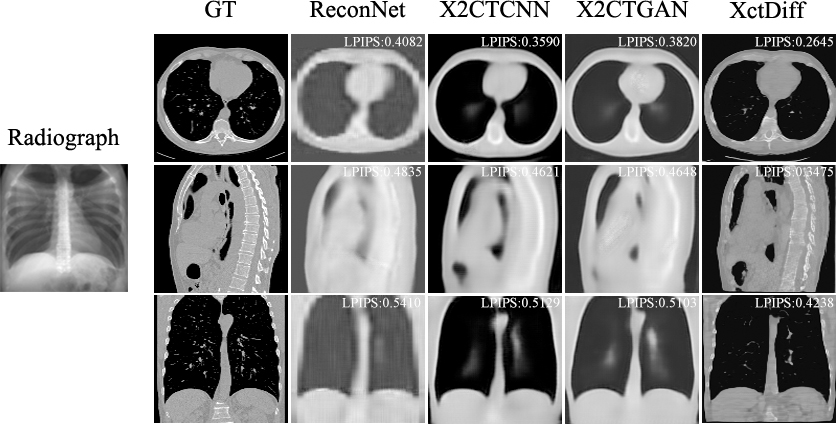}
\caption{Qualitative visualization results on the LIDC-IDRI dataset. The transverse plane, sagittal plane, and coronal plane of the reconstruction results are shown.}
\label{fig:figrue3}
\end{figure}

\subsection{Results}
\label{sec:section4.3}
\textbf{Reconstruction Result.} The quantitative results of CT reconstruction from a single radiograph are evaluated in the metrics of Peak Signal-to-Noise Ratio (PSNR), Structural Similarity (SSIM), and Learned Perceptual Image Patch Similarity (LPIPS). As shown in Table 1, it is evident that XctDiff outperforms the second-best method, X2CTGAN, by \textbf{+0.9} PSNR(dB) and \textbf{+2.1} SSIM(\%). Furthermore, as demonstrated through qualitative visualization in Fig~\ref{fig:figrue3}, our approach produces high-quality CT images with precise anatomical structures, e.g., the heart, lungs, spine, and ribs can be easily identified, which is crucial for downstream tasks. Experimental results show that CT reconstructed from a single X-ray image has much lower reconstruction accuracy than traditional CT images. But it also demonstrates the potential for certain downstream tasks.
\begin{table*}[t]
    \centering
    \renewcommand\arraystretch{1.1}
    \tabcolsep=4pt
    \caption{Ablation results on improving the structure and codebook of VQGAN.}
    \begin{tabular}{c|cc|cc|cccc|ccc}

        & \rotatebox{90}{Codebook} & \rotatebox{90}{SR Module} & \rotatebox{90}{Code Size} & \rotatebox{90}{Latent Dim} & \rotatebox{90}{PSNR $\uparrow$} & \rotatebox{90}{SSIM $\uparrow$} & \rotatebox{90}{LPIPS $\downarrow$} & \rotatebox{90}{Codebook Usage $\downarrow$}  & \rotatebox{90}{Parameters(M)} & \rotatebox{90}{FLOPs(T)} & \rotatebox{90}{GPU (G)} \\ 
        \toprule
        \multirow{2}{*}{Baseline}
        & $-$ & $-$ & 4096 & 3 & 31.95 & 0.750 & 0.060 & 1.8\% & 21.4 & 1.35 & 19.50 \\
        & $-$ & $-$ & 8192 & 8 & 32.86 & 0.769 & 0.054 & 1.5\% & 21.5 & 1.35 & 19.50 \\
        \midrule
        \multirow{3}{*}{Architecture}
        & \textcolor{red}{\checkmark} & $-$ & 4096 & 3 & 32.64 & 0.781 & 0.056 & 97.4\% & 21.4 & 1.35 & 19.50 \\
        & $-$ & \textcolor{red}{\checkmark} & 4096 & 3 & 32.23 & 0.758 & 0.056 & 1.3\% & 22.0 & 1.67 & 22.90 \\
        & \textcolor{red}{\checkmark} & \textcolor{red}{\checkmark} & 4096 & 3 & 32.74 & 0.787 & 0.049 & 98.2\% & 22.0 & 1.67 & 22.90 \\
        \midrule
        \multirow{4}{*}{Codebook}
        & \checkmark & \checkmark & 4096 & 8 & 33.35 & 0.796 & 0.048 & 96.6\% & 22.0 & 1.67 & 22.90 \\
        & \checkmark & \checkmark & 4096 & 16 & 33.01 & 0.783 & 0.051 & 55.2\% & 22.0 & 1.67 & 22.90 \\
        & \checkmark & \checkmark & 8192 & 8 & \textbf{33.43} & \textbf{0.787} & \textbf{0.049} & \textbf{98.2\%} & 22.1 & 1.67 & 22.90 \\
        & \checkmark & \checkmark & 8192 & 16 & 33.34 & 0.793 & 0.055 & 24.7\% & 22.1 & 1.67 & 22.90 \\
        \bottomrule
    \end{tabular}
\label{tab:table2}
\end{table*}
\begin{table}[b]
    \centering
    \renewcommand\arraystretch{1.1}
    \tabcolsep=5pt
    \caption{Ablation research on XctDiff framework. Note that PE and AE represent the progressive semantic encoder and improved autoencoder.}
    \begin{tabular}{c|cc|ccc}
        \toprule
        & PE & AE & PSNR $\uparrow$ & SSIM $\uparrow$ & LPIPS $\downarrow$ \\ 
        \midrule
        \multirow{4}{*}{XctDiff}
        & $-$ & $-$ & 22.47 & 0.502 & 0.154 \\
        & \textcolor{red}{\checkmark} & $-$ & 23.26 & 0.517 & 0.134 \\
        & $-$ & \textcolor{red}{\checkmark} & 23.14 & 0.511 & 0.140 \\
        & \textcolor{red}{\checkmark} & \textcolor{red}{\checkmark} & \textbf{23.57} & \textbf{0.524} & \textbf{0.123}\\
        \bottomrule
    \end{tabular}
\label{tab:table3}
\end{table}
\\
\\
\textbf{Ablation Study.} Experiment results of image quality with different settings and improvements are presented in Tab.~\ref{tab:table2}. We use two different settings as benchmarks. It can be observed that the utilization of the codebook is inefficient (only \textbf{1.8\%} and \textbf{1.5\%}, respectively), which suggests that CT images are always reconstructed based on a limited number of semantic features, resulting in a poor quality reconstruction. The third line demonstrates that the reconstructed quality is significantly improved by mapping the encoder output and codebook onto an isotropic Euclidean space. In the fourth line, the incorporation of a super-resolution module also improves the reconstruction quality. When these two components are combined, the reconstruction results surpass those achieved by the first baseline, exhibiting an improvement of \textbf{+0.8} PSNR(dB), \textbf{+3.7} SSIM(\%) and \textbf{-1.1} LPIPS(\%), respectively. In addition, we conducted an analysis on the impact of varying codebook quantities and dimensions. The results show that the best performance was achieved with a dimension set of 8 and a codebook quantity of 8192, which we use as the default setting. As shown in the Tab.~\ref{tab:table3}, we also performed ablation in the progressive semantic encoder and improved autoencoder, demonstrating their positive impact on enhancing the performance of XctDiff.
\begin{figure}[t]
\centering
\includegraphics[width=0.75\textwidth]{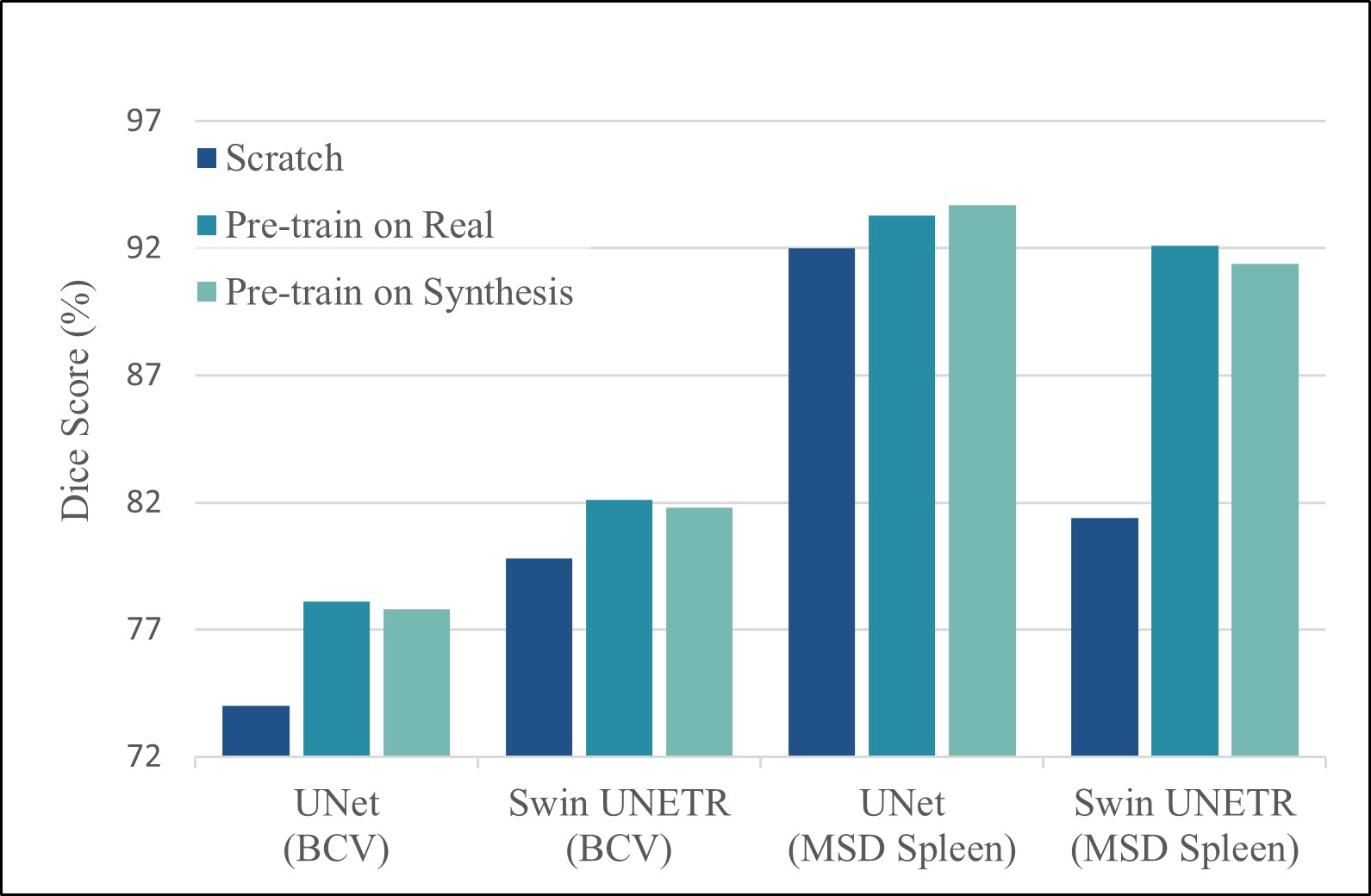}
\caption{Dice score gaps between pre-training with reconstructed CT, pre-training with real CT, and a scratch model on the BCV and MSD datasets.} 
\label{figure04}
\end{figure}
\\
\\
\textbf{Self-supervised Pre-training.} 
Reconstructed CT from a single radiograph can facilitate the self-supervised pre-training of medical images by circumventing challenges in data acquisition and privacy limitations. As shown in Fig.~\ref{figure04}, we demonstrate the efficacy of the reconstructed CT from real radiographs for the self-supervised pre-training task using two models based on different paradigms (UNet \cite{c:07} and Swin UNETR \cite{c:32}) on both BCV dataset and MSD Spleen dataset. The pre-training performance on the Reconstructed CT dataset is slightly lower than training on the real CT dataset but still with a significant improvement of \textbf{+3.1} and \textbf{+1.8} Dice(\%), respectively. For the MSD Spleen dataset, the UNet \cite{c:07} model pretrained on reconstructed CT images obtained the highest \textbf{93.7} Dice(\%), demonstrating a significant improvement of \textbf{+1.7} and \textbf{+0.4} compared to the baseline model and the model pretrained on real CT images, respectively. In addition, the Swin UNETR pre-trained on reconstructed CT dataset also exhibited competitive performance.

\section{Conclusion}
In this paper, we present an algorithmic framework XctDiff for reconstructing CT from a single radiograph, and evaluate the reconstruction performance and improvement module in detail. To facilitate real-world applications, we employ a style transfer method to convert real radiographs into synthetic-style ones for CT image reconstruction. Moreover, we explore its potential in downstream applications. The self-supervised pre-training task demonstrate the benefits of our approach in the field of medical image analysis.

\section{Acknowledgments}
The research reported in this paper is partly supported by the National Natural Science Foundation of China under grant No. 62273293, Hebei Natural Science Foundation under grant No. F2023203030, Science Research Project of Hebei Education Department under grant No. QN2024010, Shandong Provincial Natural Science Foundation under grant No. ZR2022LZH002, and partly by grant R01HL150147 from the National Institutes of Health of the United States of America.

\bibliographystyle{splncs04}
\bibliography{main}

\end{document}